\documentclass{article}
\pdfoutput=1
\usepackage{amsfonts}
\usepackage{amsmath}
\usepackage{color}
\usepackage{hyperref}
\usepackage{xcolor}
\usepackage{subcaption}
\usepackage[page]{appendix}
\usepackage{dsfont}
\usepackage{graphicx}
\DeclareMathOperator*{\esssup}{ess\,sup}
\DeclareMathOperator*{\essinf}{ess\,inf}
\newtheorem{theorem}{Theorem}

\newtheorem{corollary}[theorem]{Corollary}

\newtheorem{definition}[theorem]{Definition}
\newtheorem{example}[theorem]{Example}

\newtheorem{lemma}[theorem]{Lemma}

\newtheorem{proposition}[theorem]{Proposition}

\DeclareMathOperator*{\argmin}{arg\,min}

\begin{document}

\sloppy

\title{Optimal investment and consumption with forward preferences and uncertain parameters}
\author{W. F. Chong\thanks{%
Maxwell Institute for Mathematical Sciences and Department of Actuarial Mathematics and Statistics, Heriot-Watt University, Edinburgh EH14 4AS, U.K.; email:
\texttt{alfred.chong@hw.ac.uk}} \and G. Liang\thanks{%
Department of Statistics, The University of Warwick, Coventry CV4
7AL, U.K.;
email: \texttt{g.liang@warwick.ac.uk} }}
\date{\today}
\maketitle

\begin{abstract}
{This paper studies robust forward investment and consumption preferences within a zero-volatility context. Different from previous works, we consider an incomplete financial market model due to general investment portfolio constraints. We provide a new PDE characterization and a novel semi-explicit saddle-point construction of forward preferences and optimal strategies. We further present a more detailed construction of forward preferences and optimal strategies under constant relative risk aversion (CRRA). Key findings emphasize the necessity of a specific relationship between the initial investment preference and the forward consumption preference, indicating a long-term decreasing trend in forward consumption preference behavior.}
\end{abstract}

{\section{Introduction}
The study of continuous-time optimal investment and consumption represents an active research area in mathematical finance. In its simplest form, an agent seeks to optimize their expected inter-temporal consumption and terminal wealth preferences by selecting admissible investment and consumption strategies, denoted as $(\pi, C)$. The optimization problem takes the following form:
\begin{equation*}
\sup_{\left(\pi,C\right)}\mathbb{E}\left[\int_{0}^{T}U^c_s\left(C_s\right)ds+U_T\left(X_T^{\pi,C}\right)\right],
\end{equation*}
where $T$ is the terminal time of the investment horizon, $X^{\pi, C}$ is the derived wealth resulting from employing the investment and consumption strategies $(\pi, C)$ in a financial market model. Additionally, $U^c$ and $U$ represent \emph{static} preferences, evaluating their instantaneous consumption benefit and terminal wealth, respectively.

Therefore, a typical optimal investment and consumption model necessitates three key components: an investment horizon, a mathematical model for the financial market, and a \emph{static} performance criterion typically formulated as utility functions of consumption and wealth.

While the aforementioned optimal investment and consumption model offers valuable economic insights and maintains a close connection with stochastic control theory, it is not without limitations. Notably, the optimal investment and consumption strategies often depend on maturity, limiting the flexibility of choosing investment horizons. This leads to time-inconsistent strategies, where an optimal strategy for a time horizon $[0, T]$ may not necessarily remain optimal for $[0, S]$ with $S\neq T$, when applying the same utility functions at both $[0, T]$ and $[0, S]$.

In contrast, forward preferences, serving as a complement to static performance criteria, are constructed starting from an initial datum that represents preferences for the present rather than a future date. These preferences evolve over time, ensuring the time-consistency of optimal strategies across all time horizons. The theory, established by Musiela and Zariphopoulou in a series of works  \cite{MZ0,MZ-Kurtz,MZ-Carmona,MZ1,MZ2,MZ3}, proposes a novel approach to determining the optimal investment strategy $\pi^*$ through a \emph{forward investment preference} $U$ of wealth and time:
\begin{equation*}
U(X_t^{\pi},t)=\esssup_{\pi}\mathbb{E}\left[U\left(X_T^{\pi},T\right)|\mathcal{F}_{t}\right].
\end{equation*}
for any $0\leq t\leq T<\infty$. 
Unlike the classical framework, where future wealth preferences are assumed in advance, the forward approach requires the agent to specify their wealth preference at time $0$. Through the principles of super-martingale sub-optimality and martingale optimality (see Definition \ref{def:forward_performance_drift_vol} below, with $U^c\equiv 0$ and $\mathcal{B}$ being a singleton), the agent's forward investment preference at any time $t>0$ is then endogenously generated. This forward-looking nature enables the agent to select the optimal investment strategy $\pi^*$ without pre-defining their investment horizon and future wealth preference.

A related concept, known as horizon-unbiased utility, was introduced by Henderson in \cite{Henderson_2007} and further explored by Henderson and Hobson in \cite{Henderson_Hobson_2007}. They argue that the value functions of horizon-unbiased utilities are maturity-independent, a crucial property for optimal investment stopping problems. For a dual characterization of forward investment preference, see \cite{Zit} by {\v Z}itkovi{\'c}.
El Karoui and Mrad further provide a stochastic partial differential equation (SPDE) characterization of forward investment preference in \cite{El_Karoui_2013}. Recent developments in forward investment preference, along with applications in finance and insurance, are discussed in \cite{Angoshtari,Angoshtari2020,Anthropelos2022,Avanesyan_2018,Bo_2023,Choulli_2007,Chong_2018,Chong_2016,
El_Karoui_2018,El_Karoui_2022,El_Karoui_2021,He_2021,Hu2020,Liang2023,LZ,Nadtochiy_2017,Nadtochiy-Z,Ng_Chong_2023,Shkolnikov_2016,Strub2021} and the references therein.

In this paper, we investigate two aspects of forward preference: incorporating \emph{consumption} and \emph{parameter uncertainty}. Before presenting our contributions, we review related works. Building upon the foundation laid by Musiela and Zariphopoulou, Berrier and Tehranchi \cite{Berrier_2008} propose determining the optimal investment and consumption strategies $\left(\pi^*,C^*\right)$ for the agent through the concept of \emph{forward investment and consumption preferences} $U$ and $U^c$:
\begin{equation*}
U(X_t^{\pi},t)=\esssup_{\left(\pi,C\right)}\mathbb{E}\left[\int_{t}^{T}U^c\left(C_s,s\right)ds+U\left(X_T^{\pi,C},T\right)|\mathcal{F}_t\right].
\end{equation*}
for any $0\leq t\leq T<\infty$.
Due to the forward generating of preferences, the agent can determine their optimal investment and consumption strategies $\left(\pi^*,C^*\right)$ without pre-specifying their investment and consumption horizon, as well as their future wealth and consumption preferences. Their optimal strategies emerge as a result of the construction of the forward investment and consumption preferences (see Definition \ref{def:forward_performance_drift_vol} below, with $U^c\not\equiv0$ and $\mathcal{B}$ being a singleton).  In \cite{Berrier_2008}, the forward preferences are characterized using convex duality. In \cite{Kallblad_2016} by K{\"a}llblad, the forward preferences are characterized by a SPDE,  where zero-volatility forward preferences are linked to Black's inverse investment problem.  El Karoui et al in their work \cite{El_Karoui_2018} establish a connection between duality and the corresponding SPDE.

On the other hand, model uncertainty has been a crucial aspect in classical optimal investment and consumption problems, as demonstrated in works such as \cite{Biagini_2017, Bordigoni_2007, Fouque_2016, Hernandez_2006, Hernandez_2007, Lin_2014, Matoussi_2015, Schied_2008, Tevzadze_2013, YLZ_2017}. In reality, agents often confront ambiguity regarding financial market model uncertainty. In forward theory,
K{\"a}llblad et al in \cite{Kallblad_2018} initiated the study of the agent's robust forward investment preference (see Definition \ref{def:forward_performance_drift_vol} below, with $U^c\equiv0$ and $\mathcal{B}$ being a non-singleton) and their optimal investment strategy, focusing on the dual representation of robust forward investment preference.
However, relatively little is known about forward investment and consumption preferences with uncertainty parameters, with an exception found in \cite{Lin_2014}.

This paper makes contributions to the study of \emph{robust forward investment and consumption preferences}, with a focus on the zero-volatility case. The non-zero volatility case will be addressed in a forthcoming paper. Different from \cite{Lin_2014}, the financial market considered in this paper may be incomplete, stemming from general investment portfolio constraints. Consequently, the model includes all the previous cases studied in \cite{Berrier_2008, El_Karoui_2018, Kallblad_2016, Kallblad_2018, Lin_2014} as special cases.

As the first contribution, the paper addresses the technical challenges arising from general investment portfolio constraints in the context of financial market incompleteness. It provides a new PDE characterization and a novel semi-explicit construction of the saddle-point, which is then utilized to construct optimal investment and consumption strategies along with worst-case parameters. See Theorem \ref{prop:SPDE}. One of the major difficulties in constructing the saddle-point is establishing the continuity of the optimal investment strategy as an optimal response map to the admissible parameters, which is further compounded by the presence of general investment portfolio constraints.

Our second contribution is a more explicit construction of the forward investment and consumption preferences under the constant relative risk aversion (CRRA) assumption. Due to the homothetic assumption, the characterizing PDE reduces to an ODE. We further provide a sufficient condition for the solvability of the corresponding characterizing equation. A key finding is that to guarantee the existence of forward preferences, a parameter condition between the initial investment preference and the forward consumption preference must be satisfied (see Proposition \ref{theorem:new_theorem_1}). Essentially, it requires that the forward consumption preference must be dominated by the corresponding initial investment preference. For example, a decreasing forward consumption preference will satisfy such a requirement. In general, we show that the long-term behavior of the forward consumption preference must exhibit a decreasing trend.

This paper is structured as follows: Section 2 specifies the financial market model, incorporating general investment portfolio constraints and model uncertainty. Sections 3 and 4 introduce the robust forward investment and consumption preferences, and represent these preferences by the characterizing PDE. Section 5 constructs the CRRA-type robust forward investment and consumption preferences, solving for the robust optimal investment and consumption strategies through an ODE. The conclusion is presented in Section 6. All proofs are provided in the appendix.}

\section{The Market Model}
Let $W_t$, $ t\geq 0 $, be a $d$-dimensional Brownian motion on a probability space
$(\Omega ,\mathcal{F},\mathbb{P})$. Denote $\mathbb{F}=\{\mathcal{F}_{t}\}_{t\geq 0}$ as the augmented filtration generated by
$W$. We consider a financial market consisting of a risk-free bond, offering a constant interest rate $ r\geq 0 $, and $d$ risky stocks. The stock price processes
$S_{t}^{i}$, $t\geq 0$, solve, for each $i=1,\ldots,d$, for any $ t\geq 0 $,
\begin{equation*}
\frac{dS_{t}^{i}}{S_{t}^{i}}=b^{i}_tdt+\sigma
^{i}_tdW_{t}\text{,}
\end{equation*}%
where $ b^{i} $ and $ \sigma^{i} $ are $ \mathbb{F} $-progressively measurable processes taking values in $ \mathbb{R} $ and $ \mathbb{R}^{1\times d} $. Denote the drift vector process by $b=(b^{1},\ldots,b^{d})^{tr}$, and the volatility matrix process by $ \sigma=(\sigma^{1},\ldots,\sigma^{d})^{tr} $.

Define the set of possibly realized drift and volatility processes in the financial market as
\begin{align*}
\mathcal{B}=\{&\left(b_t,\sigma_t\right),\;t\geq 0:\left(b,\sigma\right)\text{ are }\mathbb{F}\text{-progressively measureable},\text{ and }\\&\;(b,\sigma\sigma^{tr})\in\mathbb{B}\times\Sigma,\;\mathbb{P}\times dt\text{-a.s.}\},
\end{align*}
where $\mathbb{B}$ is a convex and compact subset in $\mathbb{R}^d$, and $\Sigma$ is a convex and compact subset in $\mathcal{S}^d_{++}$, which is the set of all $d\times d$ real symmetric positive-definite matrices. Note that, for any $x_{\Sigma}\in\Sigma$, $x_{\Sigma}^{tr}=x_{\Sigma}$ and $x_{\Sigma}^{-1}$ exists.
The larger the set $\mathcal{B}$, the more uncertain the model is.	

Consider an agent who, with an initial endowment $\xi\in\mathbb{R}_{++}=\left(0,\infty\right)$, can choose to consume and invest dynamically in both the risk-free bond and risky stocks. Let $\pi_t=(\pi_t^{1},\ldots,\pi_t^{d})^{tr}$, $ t\geq 0 $, be the proportions of their wealth in the risky stocks, and let $ c_t $, $ t\geq 0 $, be their consumption rate per wealth. Then, by self-financing, their wealth process $ X_t $, $ t\geq 0 $, satisfies, for any $ t\geq 0 $,
\begin{equation}
dX_t=X_t\left(\left(r+\pi_t^{tr}\left(b_t-r\mathds{1}\right)-c_t\right)dt+\pi_t^{tr}\sigma_tdW_t\right),
\label{wealth}
\end{equation}%
with $X_0=\xi$.
Note the dependence of their wealth process $X$ on, not only their choices of the investment and consumption strategies $\left(\pi,c\right)$ and their initial endowment $\xi\in\mathbb{R}_+$, but also the market-realized drift and volatility processes $\left(b,\sigma\right)\in\mathcal{B}$ for the stock prices.
We shall occasionally write $X^{\xi;\pi,c;b,\sigma}$ for their wealth process.

Denote $\mathcal{A}_{[0,t)}$, $ t\geq 0 $, as the set of admissible investment and consumption strategies in $ [0,t) $, which is defined by, for any $ t\geq 0 $,
\begin{align*}
\mathcal{A}_{[0,t)}=\bigg\{&\left(\pi_s,c_s\right),\;s\in[0,t):\left(\pi,c\right)\text{ are }\mathbb{F}\text{-progressively measurable};\\&\left(\pi,c\right)\in\Pi\times\mathbb{R}_+,\;\mathbb{P}\times dt\text{-a.s.};\;\int_{0}^{t}\left(\vert\pi_s\vert^2+\vert c_s\vert\right)ds<\infty,\;\mathbb{P}\text{-a.s.}\bigg\},
\end{align*}
where $\Pi$ is a closed and convex subset in $\mathbb{R}^{d}$ including the origin\footnote{This assumption is in line with practice that the agent could choose to simply invest all of their wealth to the risk-free bond at any time.} $0\in\mathbb{R}^d$, and $\mathbb{R}_+=\left[0,\infty\right)$. The set of admissible investment and consumption strategies for all time $t\geq 0$ is, in turn, defined by $\mathcal{A}=\cup_{t\geq 0}\mathcal{A}_{[0,t)}$.

The above model includes the typical incomplete financial market framework, in the sense that the number of available risky stocks $ n $ could be less than the dimension $ d $ of the Brownian motion. Indeed, if the financial market only consists of $ n(<d) $ risky stocks, we could artificially construct $ d-n $ pseudo risky stocks which satisfy the assumptions above, and define the investment constraint set $ \Pi$ by $\Pi^n\times\{0\}^{d-n} $, where $ \Pi^n $ is a closed and convex subset in $ \mathbb{R}^n$ including the origin $ 0\in\mathbb{R}^n $, and $\{0\}^{d-n}$ is the subset containing only the zero vector in $ \mathbb{R}^{d-n}$. Therefore, without loss of generality, we consider the financial market model with $ n=d $.

\section{Robust Forward Investment and Consumption Preferences}
In the above financial market with model uncertainty, the agent aims to choose their optimal investment and consumption strategies $\left(\pi^*, c^*\right)\in\mathcal{A}$. Since the agent is risk-averse, their implied investment and consumption preferences are non-decreasing and concave. In particular, their time-$0$ investment and consumption preferences $U\left(x, 0\right)$ and $U^c\left(x_C, 0\right)$ are non-decreasing and concave in $x\in\mathbb{R}_{++}$ and $x_C\in\mathbb{R}_+$, where $x_C=x_cx$.

On the other hand, when determining their optimal investment and consumption strategies, the agent will consider the worst-case scenario for the average growth rate and the volatility of the stock prices in the financial market. Hence, inspired by the {\it worst-case scenario} stochastic optimization problem under the classical expected utility framework in \cite{Hernandez_2006,Hernandez_2007,Schied_2008,Tevzadze_2013,YLZ_2017}, we define the {\it robust forward investment and consumption preferences, with drift and volatility uncertainties}, and the associated optimal investment and consumption strategies, as follows.

\begin{definition} \label{def:forward_performance_drift_vol}
A pair of processes
\begin{equation*}
\{(U(\omega,x,t),U^c(\omega,x_C,t))\}_{\omega\in\Omega,x\in\mathbb{R}_{++},x_C\in\mathbb{R}_+,t\geq 0}
\end{equation*}
is called robust forward investment and consumption preferences, with drift and volatility uncertainties, if they satisfy all of the following properties:
\begin{enumerate}
\item[(i)] for each $ x\in\mathbb{R}_{++}$ and $ x_C\in\mathbb{R}_+$, $ \{U(\omega,x,t)\}_{\omega\in\Omega,t\geq 0} $ and $ \{U^c(\omega,x_C,t)\}_{\omega\in\Omega,t\geq 0} $ are $ \mathbb{F} $-progressively measurable;
\item[(ii)] for each $ \omega\in\Omega $ and $ t\geq 0 $, $ \{U(\omega,x,t)\}_{x\in\mathbb{R}_{++}} $ and $ \{U^c(\omega,x_C,t)\}_{x_C\in\mathbb{R}_+} $ are non-decreasing and concave;
\item[(iii)] for each $t\geq 0$, $\xi\in\mathcal{L}\left(\mathcal{F}_t;\mathbb{R}_{++}\right)$, and $T\geq t$,
\begin{equation}
\begin{aligned}
U\left(\xi,t\right)=\esssup_{\left(\pi,c\right)\in\mathcal{A}}\essinf_{\left(b,\sigma\right)\in\mathcal{B}}\mathbb{E}\bigg[& U\left(X_T^{\xi,t;\pi,c;b,\sigma},T\right)\\&+\int_{t}^{T}U^c\left(c_sX_s^{\xi,t;\pi,c;b,\sigma},s\right)ds\vert\mathcal{F}_t\bigg],
\end{aligned}
\label{eq:forward_self_gen}
\end{equation}
where $\mathcal{L}\left(\mathcal{F}_t;\mathbb{R}_{++}\right)$ is the set of $\mathcal{F}_t$-measurable and $\mathbb{R}_{++}$-valued random variables, and $X^{\xi,t;\pi,c;b,\sigma}$ solves \eqref{wealth} with $X^{\xi,t;\pi,c;b,\sigma}_t=\xi$.
\end{enumerate}
Moreover, if there exists a pair of forward investment and consumption strategies $\left(\pi^*,c^*\right)\in\mathcal{A}$ solving \eqref{eq:forward_self_gen} for all $t\geq 0$ and $T\geq t$, it is called optimal and robust.
\end{definition}

The pair of robust forward investment and consumption preferences include the forward investment preference, forward investment and consumption preferences, and robust forward investment preference, defined in the literature:
\begin{itemize}
\item when $U^c\equiv 0$ and $\mathcal{B}$ is a singleton, i.e., without the element of consumption and model uncertainty, the definition reduces to the forward investment preference, which was first introduced by Musiela and Zariphopoulou in a series of their works \cite{MZ0,MZ-Kurtz,MZ-Carmona,MZ1,MZ2,MZ3};
\item when $U^c\not\equiv 0$ and $\mathcal{B}$ is a singleton, i.e., with the element of consumption but without the model uncertainty, the definition reduces to the forward investment and consumption preferences, which was first introduced by Berrier and Tehranchi in \cite{Berrier_2008}, and further studied in K\"allblad \cite{Kallblad_2016} and El Karoui et al \cite{El_Karoui_2018};
\item when $U^c\equiv 0$ and $\mathcal{B}$ is not a singleton, i.e., without the element of consumption but with the model uncertainty, the definition reduces to the robust forward investment preference, which was recently introduced by K{\"a}llblad et al in \cite{Kallblad_2018}, in which they studied extensively the dual representation of robust forward investment preference. See also \cite{Lin_2020} by Lin et al.
\end{itemize}


\section{PDE Representation}
The forward investment preference, forward investment and consumption preferences, and robust forward investment preference, were characterized by SPDEs in, respectively, \cite{MZ3}, \cite{Kallblad_2016}, and \cite{Kallblad_2018}, with the novel element of volatility processes as a model input. Instead of endogenously being determined by the dynamic programming principle as in the classical framework, the volatility of forward preferences is exogenously chosen by the agent, and is regarded as their belief on how their preference in the performance criterion is going to evolve in the future. That is, in general, $U\left(x,t\right)$, $x\in\mathbb{R}_{++}$ and $t\geq 0$, in Definition \ref{def:forward_performance_drift_vol} admits an It{\^o}'s decomposition: for any $x\in\mathbb{R}_{++}$ and $t\geq 0$,
\begin{equation*}
dU(x,t)=f\left(x,t\right)dt+a(x,t)^{tr}dW_t,
\end{equation*}
for some $\mathbb{F}$-progressively measurable processes $f\left(x,t\right)$ and $a\left(x,t\right)$, $x\in\mathbb{R}_{++}$ and $t\geq 0$, taking values in $\mathbb{R}$ and $\mathbb{R}^d$ respectively. The process $a\left(x,t\right)$ represents the volatility of forward preferences and is to be chosen as a model input.

In this paper, we assume that the agent adapts a {\it zero-volatility} robust forward preference, with $a\left(x,t\right)\equiv 0$ for any $x\in\mathbb{R}_{++}$ and $t\geq 0$.
The non-zero volatility case will be studied in another paper \cite{CL2023}.

Due to the zero-volatility assumption, the SPDE will reduce to a PDE, and this section provides a PDE representation for the zero-volatility robust forward investment and consumption preferences, with drift and volatility uncertainties. To this end, let $U\left(x,t\right)$ and $U^c\left(x_C,t\right)$, $x\in\mathbb{R}_{++}$, $x_C\in\mathbb{R}_+$, and $t\geq 0$, be a pair of deterministic functions, such that, for any $t\geq 0$, $U\left(x,t\right)$ and $U^c\left(x_C,t\right)$ are non-decreasing and concave, in $x\in\mathbb{R}_{++}$ and $x_C\in\mathbb{R}_+$ respectively, and $U\left(\cdot,\cdot\right),U^c\left(\cdot,\cdot\right)\in C^{2,1}$, which is the class of all twice differentiable, with respect to the first argument, and differentiable, with respect to the second argument, functions with continuous partial derivatives.

We first present a key lemma that provides a candidate saddle-point for the optimal investment and consumption strategies and the worst-case drift and volatility processes.
It is noteworthy that while the classical Sion's minmax theorem confirms the existence of saddle-points, it does not provide how to construct them. However, leveraging the specific structure of the function $F$ below, we can construct its saddle-point in a semi-explicit form, going beyond merely demonstrating its existence.
A major technical difficulty is to establish the continuity of the optimal investment strategy as an optimal
response map to the admissible parameters, which is further compounded by the presence of general
investment portfolio constraints. The proof of this aspect is detailed in step 2 of the following lemma.

\begin{lemma}\label{lemma:saddle_point}
For each $x\in\mathbb{R}_{++}$ and $t\geq 0$, the deterministic function, for any $\left(x_{\pi};x_{b},x_{\Sigma}\right)\in\Pi\times\mathbb{B}\times\Sigma$,
\begin{equation}
F\left(x,t;x_{\pi};x_{b},x_{\Sigma}\right)=\frac{1}{2}x^2U_{xx}\left(x,t\right)x_{\pi}^{tr}x_{\Sigma}x_{\pi}+xU_x\left(x,t\right)x_{\pi}^{tr}\left(x_{b}-r\mathds{1}\right),
\label{eq:F}
\end{equation}
admits a saddle-point $\left(x^*_{\pi};x^*_{b},x^*_{\Sigma}\right)\in\Pi\times\mathbb{B}\times\Sigma$ given in (\ref{minimizer}) and (\ref{maximizer}) in the appendix, which depends on $\left(x,t\right)\in \mathbb{R}_{++}\times\left[0,\infty\right)$, in the sense that, for any $x_{\pi}\in\Pi$ and $\left(x_{b},x_{\Sigma}\right)\in\mathbb{B}\times\Sigma$,
\begin{equation}
F\left(x,t;x_{\pi};x^*_{b},x^*_{\Sigma}\right)\leq F\left(x,t;x^*_{\pi};x^*_{b},x^*_{\Sigma}\right)\leq F\left(x,t;x^*_{\pi};x_{b},x_{\Sigma}\right).
\label{eq:saddle_inequality}
\end{equation}
\end{lemma}

Define, for any $t\geq 0$,
\begin{equation}
\pi^*_t=x^*_{\pi}\left(X_t,t\right)\quad\text{and}\quad c^*_t=x^*_{c}\left(X_t,t\right).
\label{eq:optimal_strategies}
\end{equation}
Herein, $x^*_{\pi}\in\Pi$ such that $\left(x^*_{\pi};x^*_{b},x^*_{\Sigma}\right)\in\Pi\times\mathbb{B}\times\Sigma$, which depends on $\left(x,t\right)\in \mathbb{R}_{++}\times\left[0,\infty\right)$, is a saddle-point in Lemma \ref{lemma:saddle_point}; in addition, $x^*_c\in\mathbb{R}_+$, which also depends on $\left(x,t\right)\in \mathbb{R}_{++}\times\left[0,\infty\right)$, is given by
\begin{equation*}
x^*_c=
\begin{cases}
0&\text{if }\lim_{x_c\rightarrow 0+}U^c_{x_C}\left(x_cx,t\right)\leq U_x\left(x,t\right),\\
\tilde{x}_c&\text{if }\lim_{x_c\rightarrow\infty}U^c_{x_C}\left(x_cx,t\right)<U_x\left(x,t\right)<\lim_{x_c\rightarrow 0+}U^c_{x_C}\left(x_cx,t\right)\\
\infty&\text{if }U_x\left(x,t\right)\leq\lim_{x_c\rightarrow\infty}U^c_{x_C}\left(x_cx,t\right),
\end{cases},
\end{equation*}
where $\tilde{x}_c\in\mathbb{R}_{++}$ solves the equation $U^c_{x_C}\left(\tilde{x}_cx,t\right)=U_x\left(x,t\right)$.\\

The main result of this paper is the following verification theorem. 

\begin{theorem}\label{prop:SPDE}
Suppose that $a\left(x,t\right)\equiv 0$ for any $x\in\mathbb{R}_{++}$ and $t\geq 0$, so that, for any $x\in\mathbb{R}_{++}$ and $t\geq 0$,
\begin{equation}
dU\left(x,t\right)=f\left(x,t\right)dt,
\label{eq:PDE}
\end{equation}
for some deterministic function $f\left(x,t\right)$, $x\in\mathbb{R}_{++}$ and $t\geq 0$. Suppose further that the deterministic functions $U\left(x,t\right)$ and $U^c\left(x_C,t\right)$, $x\in\mathbb{R}_{++}$, $x_C\in\mathbb{R}_+$, and $t\geq 0$, satisfy the following conditions:
\begin{enumerate}
\item[(i)] there exists an $M_1\in\mathbb{R}$ such that, for any $x\in\mathbb{R}_{++}$ and $t\geq 0$, $U\left(x,t\right)\geq M_1$;
\item[(ii)] for any $\left(b,\sigma\right)\in\mathcal{B}$ and $t\geq 0$,
\begin{equation*}
\left\{U\left(X^{\xi;\pi^*,c^*;b,\sigma}_{\tau},\tau\right)\right\}_{\tau\in\mathcal{T}\left[0,t\right]}
\end{equation*}
is uniformly integrable, where $\mathcal{T}\left[0,t\right]$ is the set of all $\mathbb{F}$-stopping time $\tau\in\left[0,t\right]$;
\item[(iii)] there exists an $M_2\in\mathbb{R}$ such that $\int_{0}^{\infty}\min\left\{U^c\left(0,s\right),0\right\}ds\geq M_2$.
\end{enumerate}
Then the pair of deterministic functions $U\left(x,t\right)$ and $U^c\left(x_C,t\right)$, $x\in\mathbb{R}_{++}$, $x_C\in\mathbb{R}_+$, and $t\geq 0$, are zero-volatility robust forward investment and consumption preferences, with drift and volatility uncertainties, if and only if, for any $x\in\mathbb{R}_{++}$ and $t\geq 0$,
\begin{equation}
\begin{aligned}
f\left(x,t\right)=&-F\left(x,t;x^*_{\pi};x^*_{b},x^*_{\Sigma}\right)-\sup_{x_c\in\mathbb{R}_+}\left(U^c\left(x_cx,t\right)-x_cxU_x\left(x,t\right)\right)\\&-rxU_x\left(x,t\right).
\end{aligned}
\label{eq:equality}
\end{equation}
Moreover, if, for any $t\geq 0$, $\int_{0}^{t}\left(\vert\pi^*_s\vert^2+\vert c^*_s\vert\right)ds<\infty,\;\mathbb{P}\text{-a.s.}$, the optimal and robust forward investment and consumption strategies are given by $\pi^*$ and $c^*$ in \eqref{eq:optimal_strategies}.
\end{theorem}

\section{Homothetic Robust Forward CRRA Preferences}
In this section, by making use of Theorem \ref{prop:SPDE}, we construct homothetic zero-volatility robust forward investment and consumption CRRA preferences, with drift and volatility uncertainties, from the solutions of a family of ODEs. The motivation behind constructing the homothetic forward CRRA preferences is that, in the classical framework, the value functions, also known as backward preferences, are homothetic. This can be observed in, for instance, Hu et al \cite{HU0} and Cheridito and Hu \cite{Cheridito_2011}, where the SPDE representation reduces to a BSDE representation for the value functions. In the forward framework with the zero-volatility assumption, the PDE representation will reduce to an ODE representation.

Suppose that the deterministic functions take the following forms: for any $x\in\mathbb{R}_{++}$, $x_C\in\mathbb{R}_+$, and $t\geq 0$,
\begin{equation}
U(x,t)=\frac{x^{\delta}}{\delta}e^{Y_t}\quad\text{and}\quad U^c(x_C,t)=\frac{x_C^{\delta}}{\delta}\lambda_t,
\label{eq:forward_utility_drift_vol_uncertainty}
\end{equation}
where $\delta\in(0,1)$ is the agent's risk aversion parameter constant, and where the homothetic factors $e^{Y_t}$ and $\lambda_t$ are two deterministic functions in time $t\geq 0$ which are independent of the state variables $x\in\mathbb{R}_{++}$, $x_C\in\mathbb{R}_+$. The homothetic factor of $U$ being in an exponential form is inspired by the classical framework, as seen in, for instance, \cite{HU0} and \cite{Cheridito_2011}. The homothetic factor of $U^c$ is generic in this context and will be discussed in detail after the following proposition on constructing the agent's preferences.

\begin{proposition}\label{theorem:new_theorem_1}
Let $Y_t$, $t\geq 0$, be a deterministic function which is differentiable with a continuous derivative, and let $\lambda_t$, $t\geq 0$, be another deterministic function which is non-negative, bounded, and differentiable with a continuous derivative. Assume that $Y$ satisfies the following ODE, while $\lambda$ satisfies the following integrability condition: for any $t\geq 0$,
\begin{equation}
dY_t=-\left(G+(1-\delta)\lambda_t^{\frac{1}{1-\delta}}e^{-\frac{Y_t}{1-\delta}}+\delta r\right)dt,
\label{eq:new_ODE}
\end{equation}
\begin{equation}
e^{\frac{Y_0}{1-\delta}}>\int_{0}^{\infty}e^{\frac{G+\delta r}{1-\delta}s}\lambda_s^{\frac{1}{1-\delta}}ds.
\label{eq:condition_1}
\end{equation}
Herein, $G\geq 0$ is a saddle-value of the deterministic function: for any $\left(x_{\pi};x_{b},x_{\Sigma}\right)\in\Pi\times\mathbb{B}\times\Sigma$,
\begin{equation}
G\left(x_{\pi};x_{b},x_{\Sigma}\right)=\frac{1}{2}\delta\left(\delta-1\right)x_{\pi}^{tr}x_{\Sigma}x_{\pi}+\delta x_{\pi}^{tr}\left(x_{b}-r\mathds{1}\right);
\label{eq:function_G}
\end{equation}
that is, $G=G\left(x^*_{\pi};x^*_{b},x^*_{\Sigma}\right)$ is independent of $\left(x,t\right)\in \mathbb{R}_{++}\times\left[0,\infty\right)$, with a saddle-point $\left(x^*_{\pi};x^*_{b},x^*_{\Sigma}\right)\in\Pi\times\mathbb{B}\times\Sigma$, which is also independent of $\left(x,t\right)\in \mathbb{R}_{++}\times\left[0,\infty\right)$, satisfying that, for any $x_{\pi}\in\Pi$ and $\left(x_{b},x_{\Sigma}\right)\in\mathbb{B}\times\Sigma$,
\begin{equation*}
G\left(x_{\pi};x^*_{b},x^*_{\Sigma}\right)\leq G\left(x^*_{\pi};x^*_{b},x^*_{\Sigma}\right)\leq G\left(x^*_{\pi};x_{b},x_{\Sigma}\right).
\end{equation*}
Then the pair of deterministic functions $U\left(x,t\right)$ and $U^c\left(x_C,t\right)$, $x\in\mathbb{R}_{++}$, $x_C\in\mathbb{R}_+$, and $t\geq 0$, given in \eqref{eq:forward_utility_drift_vol_uncertainty} are zero-volatility robust forward investment and consumption CRRA preferences, with drift and volatility uncertainties. Moreover, the optimal and robust forward investment and consumption strategies are given by $\left(\pi^*,c^*\right)\in\mathcal{A}$, which are defined by, for any $t\geq 0$,
\begin{equation}
\pi^*_t=x^*_{\pi}\quad\text{and}\quad c^*_t=\lambda_t^{\frac{1}{1-\delta}}e^{-\frac{Y_t}{1-\delta}}.
\label{eq:optimal_robust_strategy}
\end{equation}
\end{proposition}

\subsection{Homothetic Factor of Consumption Preference}

Condition \eqref{eq:condition_1} on the relation between $Y_0$ and $\lambda$ might seem to be stringent at the first glance; yet, a rich class of non-negative, bounded, and differentiable with a continuous derivative, deterministic functions $\lambda$ actually satisfies this condition. For example, suppose that the function $\lambda$ is given by, for any $t\geq 0$,
\begin{equation*}
\lambda_t=\alpha e^{-\left(G+\delta r+\beta\right)t},
\end{equation*}
for some constants $\alpha\geq 0$ and $\beta>0$. Sufficiently, condition (\ref{eq:condition_1}) is satisfied when
\begin{equation}
e^{\frac{Y_0}{1-\delta}}>\frac{1-\delta}{\beta}\alpha^{\frac{1}{1-\delta}}.
\label{eq:sufficient_condition}
\end{equation}
With this class of deterministic functions $\lambda$, by Proposition \ref{theorem:new_theorem_1}, the zero-volatility robust forward investment and consumption CRRA preferences, with drift and volatility uncertainties, are given by, for any $x\in\mathbb{R}_{++}$, $x_C\in\mathbb{R}_+$, and $t\geq 0$,
\begin{equation*}
U\left(x,t\right)=\frac{x^{\delta}}{\delta}e^{-\left(G+\delta r\right)t}\left(\frac{1-\delta}{\beta}\alpha^{\frac{1}{1-\delta}}e^{-\frac{\beta}{1-\delta}t}+e^{\frac{Y_0}{1-\delta}}-\frac{1-\delta}{\beta}\alpha^{\frac{1}{1-\delta}}\right)^{1-\delta}
\end{equation*}
and $U^c\left(x_C,t\right)=\frac{x_C^{\delta}}{\delta}\alpha e^{-\left(G+\delta r+\beta\right)t}$, with the corresponding optimal and robust forward consumption strategy given by, for any $t\geq 0$,
\begin{equation*}
c^*_t=\frac{\alpha^{\frac{1}{1-\delta}}e^{-\frac{\beta}{1-\delta}t}}{\frac{1-\delta}{\beta}\alpha^{\frac{1}{1-\delta}}e^{-\frac{\beta}{1-\delta}t}+e^{\frac{Y_0}{1-\delta}}-\frac{1-\delta}{\beta}\alpha^{\frac{1}{1-\delta}}},
\end{equation*}
which is clearly non-increasing and converges to $0$ when time $t\rightarrow\infty$.

To satisfy the sufficient condition \eqref{eq:sufficient_condition}, an obvious instance is putting $\alpha=0$, which corresponds to the case when $\lambda\equiv 0$.
\begin{corollary}\label{coro1}
Let $G=G\left(x^*_{\pi};x^*_{b},x^*_{\Sigma}\right)\geq 0$ and $\left(x^*_{\pi};x^*_{b},x^*_{\Sigma}\right)\in\Pi\times\mathbb{B}\times\Sigma$ be a saddle-value and a saddle-point of the deterministic function \eqref{eq:function_G}. The deterministic function $U\left(x,t\right)=\frac{x^{\delta}}{\delta}e^{Y_0-\left(G+\delta r\right)t}$, $x\in\mathbb{R}_{++}$ and $t\geq 0$, is the zero-volatility robust forward investment CRRA preference, with drift and volatility uncertainties. Moreover, the optimal and robust forward investment strategy is given by $\pi^*\in\mathcal{A}$, which is defined by, for any $t\geq 0$, $\pi^*_t=x^*_{\pi}$.
\end{corollary}

Another possible choice to satisfy the sufficient condition (\ref{eq:sufficient_condition}) is when (i) $\alpha=e^{Y_0}$, so that the initial investment and consumption preferences of the agent coincide, and (ii) $\beta>1-\delta$.
\begin{corollary}\label{coro2}
Let $G=G\left(x^*_{\pi};x^*_{b},x^*_{\Sigma}\right)\geq 0$ and $\left(x^*_{\pi};x^*_{b},x^*_{\Sigma}\right)\in\Pi\times\mathbb{B}\times\Sigma$ be a saddle-value and a saddle-point of the deterministic function \eqref{eq:function_G}. The pair of deterministic functions, for any $x\in\mathbb{R}_{++}$, $x_C\in\mathbb{R}_+$, and $t\geq 0$,
\begin{equation*}
U\left(x,t\right)=\frac{x^{\delta}}{\delta}e^{Y_0-\left(G+\delta r\right)t}\left(\frac{1-\delta}{\beta}e^{-\frac{\beta}{1-\delta}t}+1-\frac{1-\delta}{\beta}\right)^{1-\delta}
\end{equation*}
and $U^c\left(x_C,t\right)=\frac{x_C^{\delta}}{\delta}e^{Y_0-\left(G+\delta r\right)t}e^{-\beta t}$, is the zero-volatility robust forward investment and consumption CRRA preferences, with drift and volatility uncertainties. Moreover, the optimal and robust forward investment and consumption strategies are given by $\left(\pi^*,c^*\right)$, which are defined by, for any $t\geq 0$,
\begin{equation*}
\pi^*_t=x^*_{\pi}\quad\text{and}\quad c^*_t=\frac{e^{-\frac{\beta}{1-\delta}t}}{\frac{1-\delta}{\beta}e^{-\frac{\beta}{1-\delta}t}+1-\frac{1-\delta}{\beta}}.
\end{equation*}
\end{corollary}

\subsection{Time Monotonicity}
Notice that, in Proposition \ref{theorem:new_theorem_1}, the deterministic function $Y_t$, $t\geq 0$, is non-increasing. Therefore, the constructed zero-volatility robust forward investment CRRA preference, with drift and volatility uncertainties, given in \eqref{eq:forward_utility_drift_vol_uncertainty}, is time-monotonic, more precisely non-increasing. Such a time monotonicity of the robust forward investment preference is not due to the consumption component; see, for instance, Corollary \ref{coro1}. This result coincides with, for example, \cite{Kallblad_2018}, \cite{LZ}, \cite{MZ0} and \cite{MZ2}, in which the zero-volatility forward investment preference is shown to be non-increasing in time.

On the other hand, the deterministic function $\lambda_t$, $t\geq 0$, and hence the constructed zero-volatility robust forward consumption CRRA preference, with drift and volatility uncertainties, given in \eqref{eq:forward_utility_drift_vol_uncertainty}, are not necessarily non-decreasing nor non-increasing at all time $t\geq 0$. However, they are necessarily non-increasing for all time $t\geq t^*$, for some threshold time $t^*\geq 0$, in order to satisfy the condition \eqref{eq:condition_1} for all time $t\geq 0$. In Corollary \ref{coro2}, such an asymptotic time-monotonicity is clear, and the zero-volatility robust forward investment and consumption CRRA preferences, with drift and volatility uncertainties, asymptotically converge at the same rate that, for any $x\in\mathbb{R}_{++}$ and $x_C\in\mathbb{R}_+$,
\begin{equation*}
\lim_{t\rightarrow\infty}\frac{\frac{U\left(x,t\right)}{x^{\delta}}}{\frac{U^c\left(x_C,t\right)}{x_C^{\delta}}}=1.
\end{equation*}

\subsection{Example with Explicit Saddle-Point and Saddle-Value}
In Proposition \ref{theorem:new_theorem_1}, and consequently in Corollaries \ref{coro1} and \ref{coro2}, the constructed zero-volatility robust forward investment and consumption CRRA preferences, with drift and volatility uncertainties, and the corresponding optimal and robust forward investment and consumption strategies, depend on the saddle-point and the saddle-value of the deterministic function in \eqref{eq:function_G}. We demonstrate an example with an explicit saddle-point and saddle-value as follows.

\begin{example}
Consider the case that $d=1$, and let $\mathbb{B}=\left[\underline{b},\overline{b}\right]$, $\Sigma=\left[\underline{\Sigma},\overline{\Sigma}\right]$, and $\Pi=\left[\underline{\pi},\overline{\pi}\right]$, where $\underline{b},\overline{b}\in\mathbb{R}$, $\underline{\Sigma},\overline{\Sigma}\in\mathbb{R}_{++}$, and $\underline{\pi},\overline{\pi}\in\overline{\mathbb{R}}=\left[-\infty,\infty\right]$ with $\underline{\pi}\leq 0\leq\overline{\pi}$. The saddle-point $\left(x^*_{\pi};x^*_{b},x^*_{\Sigma}\right)\in\Pi\times\mathbb{B}\times\Sigma$ of the deterministic function in \eqref{eq:function_G} is given by
\begin{equation}
x^*_{\pi}=\min\left\{\overline{\pi},\frac{\underline{b}-r}{\left(1-\delta\right)\overline{\Sigma}}\right\}\mathds{1}_{\left\{r\leq\underline{b}\right\}}+\max\left\{\underline{\pi},\frac{\overline{b}-r}{(1-\delta)\overline{\Sigma}}\right\}\mathds{1}_{\left\{r\geq\overline{b}\right\}};
\label{eq:pi_example}
\end{equation}
\begin{equation*}
x^*_{b}=\underline{b}\mathds{1}_{\left\{r\leq\underline{b}\right\}}+r\mathds{1}_{\left\{\underline{b}<r<\overline{b}\right\}}+\overline{b}\mathds{1}_{\left\{r\geq\overline{b}\right\}}\quad\text{and}\quad x^*_{\Sigma}=\overline{\Sigma};
\end{equation*}
its saddle-value $G\geq 0$ is given by
\begin{equation}
G=\frac{1}{2}\delta\left(\delta-1\right)\overline{\Sigma}\text{dist}^2\left\{\left[\underline{\pi},\overline{\pi}\right],\frac{x^*_b-r}{\left(1-\delta\right)\overline{\Sigma}}\right\}+\frac{1}{2}\frac{\delta}{1-\delta}\frac{\left(x^*_b-r\right)^2}{\overline{\Sigma}}.
\label{eq:G_example}
\end{equation}
Then, in Proposition \ref{theorem:new_theorem_1}, the zero-volatility robust forward investment and consumption CRRA preferences, with drift and volatility uncertainties, are with $G\geq 0$ given in \eqref{eq:G_example}; the corresponding optimal and robust forward investment and consumption strategies are with $x^*_{\pi}\in\Pi$ and $G\geq 0$ given in \eqref{eq:pi_example} and \eqref{eq:optimal_robust_strategy}. In particular, if the risk-free rate $r$ lies between the lowest possible drift $\underline{b}$ and the greatest possible drift $\overline{b}$, the agent should optimally invest all of their wealth into the risk-free bond at all future time $t\geq 0$.
\end{example}

\section{Concluding Remarks and Extensions}

{In this paper, we have solved the robust optimal investment and consumption strategies for an agent employing robust forward preferences. We have characterized these robust forward investment and consumption preferences for the agent using a PDE, in the presence of investment constraints. Furthermore, 
using the element of ODE, we have explicitly constructed the robust CRRA forward preferences and the associated optimal strategies. An interesting finding is that the forward consumption preference must be dominated by the initial investment preference, leading to a noticeable long-term decreasing trend in the forward consumption preference.

The paper focuses on the zero-volatility case. The non-zero volatility is far more challenging. It is expected that the corresponding characterizing PDE will transform to a fully nonlinear SPDE. In the CRRA case, the ODE will become a backward stochastic differential equation (BSDE). However, it is worth noting that the corresponding saddle point in the non-zero volatility case doesn't generally exist. To address this challenge, we will propose a randomization approach that expands the space of uncertainty parameters to the probability distributions of these parameters. This will lead to a new class of BSDEs. Our forthcoming paper \cite{CL2023} will discuss this non-zero volatility case and its corresponding BSDE characterization.}

\renewcommand*\appendixpagename{\Large Appendices}
\begin{appendices}
\section{Proof of Lemma \ref{lemma:saddle_point}}

\underline{Step 1}. 
Let $x\in\mathbb{R}_{++}$ and $t\geq 0$. Note that the deterministic function in \eqref{eq:F} can be rewritten as, for any $x_{\pi}\in\Pi$ and $\left(x_{b},x_{\Sigma}\right)\in\mathbb{B}\times\Sigma$,
\begin{align*}
&\;F\left(x,t;x_{\pi};x_{b},x_{\Sigma}\right)\\=&\;\frac{1}{2}x^2U_{xx}\left(x,t\right)\left(x_{\pi}+\frac{U_{x}\left(x,t\right)}{xU_{xx}\left(x,t\right)}x_{\Sigma}^{-1}\left(x_{b}-r\mathds{1}\right)\right)^{tr}\\&\;x_{\Sigma}\left(x_{\pi}+\frac{U_{x}\left(x,t\right)}{xU_{xx}\left(x,t\right)}x_{\Sigma}^{-1}\left(x_{b}-r\mathds{1}\right)\right)-\frac{1}{2}\frac{U_{x}\left(x,t\right)^2}{U_{xx}\left(x,t\right)}\left(x_{b}-r\mathds{1}\right)^{tr}x_{\Sigma}^{-1}\left(x_{b}-r\mathds{1}\right).
\end{align*}

Fix any $\left(x_{b},x_{\Sigma}\right)\in\mathbb{B}\times\Sigma$. Then,
\begin{equation}
\begin{aligned}
&\;\tilde{x}_{\pi}\left(x_{b},x_{\Sigma}\right)\\=&\;\argmin_{x_{\pi}\in\Pi}\left(\left(x_{\pi}+\frac{U_{x}\left(x,t\right)}{xU_{xx}\left(x,t\right)}x_{\Sigma}^{-1}\left(x_{b}-r\mathds{1}\right)\right)^{tr}\right.\\&\;\left.\quad\quad\quad\quad\;\; x_{\Sigma}\left(x_{\pi}+\frac{U_{x}\left(x,t\right)}{xU_{xx}\left(x,t\right)}x_{\Sigma}^{-1}\left(x_{b}-r\mathds{1}\right)\right)\right)\\=&\;-\frac{U_{x}\left(x,t\right)}{xU_{xx}\left(x,t\right)}x_{\Sigma}^{-1}\left(x_{b}-r\mathds{1}\right)\mathds{1}_{\left\{-\frac{U_{x}\left(x,t\right)}{xU_{xx}\left(x,t\right)}x_{\Sigma}^{-1}\left(x_{b}-r\mathds{1}\right)\in\Pi\right\}}\\&\;+\argmin_{x_{\pi}\in\partial\Pi}\left(\left(x_{\pi}+\frac{U_{x}\left(x,t\right)}{xU_{xx}\left(x,t\right)}x_{\Sigma}^{-1}\left(x_{b}-r\mathds{1}\right)\right)^{tr}\right.\\&\;\left.\quad\quad\quad\quad\quad\;\; x_{\Sigma}\left(x_{\pi}+\frac{U_{x}\left(x,t\right)}{xU_{xx}\left(x,t\right)}x_{\Sigma}^{-1}\left(x_{b}-r\mathds{1}\right)\right)\right)\mathds{1}_{\left\{-\frac{U_{x}\left(x,t\right)}{xU_{xx}\left(x,t\right)}x_{\Sigma}^{-1}\left(x_{b}-r\mathds{1}\right)\notin\Pi\right\}}
\end{aligned}
\label{eq:tilde_pi}
\end{equation}
maximizes \eqref{eq:F}, where $\partial\Pi$ is the boundary of the closed and convex set $\Pi$. If $-\frac{U_{x}\left(x,t\right)}{xU_{xx}\left(x,t\right)}x_{\Sigma}^{-1}\left(x_{b}-r\mathds{1}\right)\in\Pi$, the construction is obvious. Suppose that $-\frac{U_{x}\left(x,t\right)}{xU_{xx}\left(x,t\right)}x_{\Sigma}^{-1}\left(x_{b}-r\mathds{1}\right)\notin\Pi$, and assume that there exists an $x_{\pi}'\left(x_{b},x_{\Sigma}\right)\in\Pi^{\circ}$, which is the interior of the set $\Pi$, which minimizes
\begin{align*}
&\;Q\left(x,t;x_{\pi};x_{b},x_{\Sigma}\right)\\=&\;\left(x_{\pi}+\frac{U_{x}\left(x,t\right)}{xU_{xx}\left(x,t\right)}x_{\Sigma}^{-1}\left(x_{b}-r\mathds{1}\right)\right)^{tr}x_{\Sigma}\left(x_{\pi}+\frac{U_{x}\left(x,t\right)}{xU_{xx}\left(x,t\right)}x_{\Sigma}^{-1}\left(x_{b}-r\mathds{1}\right)\right).
\end{align*}
That is, for any $x_{\pi}\in\Pi$,
\begin{equation}
0<Q\left(x,t;x_{\pi}'\left(x_{b},x_{\Sigma}\right);x_{b},x_{\Sigma}\right)\leq Q\left(x,t;x_{\pi};x_{b},x_{\Sigma}\right),
\label{eq:Q_contradiction}
\end{equation}
where the first strict inequality is due to the facts that $x_{\Sigma}$ is positive-definite and $x_{\pi}'\left(x_{b},x_{\Sigma}\right)\neq-\frac{U_{x}\left(x,t\right)}{xU_{xx}\left(x,t\right)}x_{\Sigma}^{-1}\left(x_{b}-r\mathds{1}\right)$. Since $x_{\pi}'\left(x_{b},x_{\Sigma}\right)\in\Pi^{\circ}$, there exists an $\theta\in\left(0,1\right)$ being close enough to $1$ such that
\begin{equation*}
\bar{x}_{\pi}\left(x_{b},x_{\Sigma}\right)=\theta x_{\pi}'\left(x_{b},x_{\Sigma}\right)+\left(1-\theta\right)\left(-\frac{U_{x}\left(x,t\right)}{xU_{xx}\left(x,t\right)}x_{\Sigma}^{-1}\left(x_{b}-r\mathds{1}\right)\right)\in\Pi^{\circ}\subseteq\Pi.
\end{equation*}
By the positive-definiteness of $x_{\Sigma}$, and thus the strict convexity of $Q\left(x,t;\cdot;x_{b},x_{\Sigma}\right)$,
\begin{align*}
&\;Q\left(x,t;\bar{x}_{\pi}\left(x_{b},x_{\Sigma}\right);x_{b},x_{\Sigma}\right)\\<&\;\theta Q\left(x,t;x_{\pi}'\left(x_{b},x_{\Sigma}\right);x_{b},x_{\Sigma}\right)\\&\;+\left(1-\theta\right)Q\left(x,t;-\frac{U_{x}\left(x,t\right)}{xU_{xx}\left(x,t\right)}x_{\Sigma}^{-1}\left(x_{b}-r\mathds{1}\right);x_{b},x_{\Sigma}\right)\\=&\;\theta Q\left(x,t;x_{\pi}'\left(x_{b},x_{\Sigma}\right);x_{b},x_{\Sigma}\right)\\<&\;Q\left(x,t;x_{\pi}'\left(x_{b},x_{\Sigma}\right);x_{b},x_{\Sigma}\right),
\end{align*}
which contradicts \eqref{eq:Q_contradiction}.

\underline{Step 2}.
Next, we prove that any maximizer $\tilde{x}_{\pi}\left(x_{b},x_{\Sigma}\right)\in\Pi$ of \eqref{eq:F}, in particular \eqref{eq:tilde_pi}, is continuous in $\left(x_{b},x_{\Sigma}\right)\in\mathbb{B}\times\Sigma$. Let $\left(x_{b},x_{\Sigma}\right),\left(x'_{b},x'_{\Sigma}\right)\in\mathbb{B}\times\Sigma$. By the convexity of $\Pi$, for any $\eta\in\left(0,1\right]$ and $\hat{x}_{\pi}\in\Pi$, $\left(1-\eta\right)\tilde{x}_{\pi}\left(x_{b},x_{\Sigma}\right)+\eta\hat{x}_{\pi}\in\Pi$, and thus
\begin{align*}
&\;F\left(x,t;\tilde{x}_{\pi}\left(x_{b},x_{\Sigma}\right);x_{b},x_{\Sigma}\right)\\\geq&\;F\left(x,t;\left(1-\eta\right)\tilde{x}_{\pi}\left(x_{b},x_{\Sigma}\right)+\eta\hat{x}_{\pi};x_{b},x_{\Sigma}\right)\\=&\;\frac{1}{2}x^2U_{xx}\left(x,t\right)\left(\left(1-\eta\right)\tilde{x}_{\pi}\left(x_{b},x_{\Sigma}\right)+\eta\hat{x}_{\pi}\right)^{tr}x_{\Sigma}\left(\left(1-\eta\right)\tilde{x}_{\pi}\left(x_{b},x_{\Sigma}\right)+\eta\hat{x}_{\pi}\right)\\&\;+xU_x\left(x,t\right)\left(\left(1-\eta\right)\tilde{x}_{\pi}\left(x_{b},x_{\Sigma}\right)+\eta\hat{x}_{\pi}\right)^{tr}\left(x_{b}-r\mathds{1}\right)\\=&\;\frac{1}{2}x^2U_{xx}\left(x,t\right)\left(\tilde{x}_{\pi}\left(x_{b},x_{\Sigma}\right)+\eta\left(\hat{x}_{\pi}-\tilde{x}_{\pi}\left(x_{b},x_{\Sigma}\right)\right)\right)^{tr}\\&\;\quad\quad\quad\quad\quad\quad\quad x_{\Sigma}\left(\tilde{x}_{\pi}\left(x_{b},x_{\Sigma}\right)+\eta\left(\hat{x}_{\pi}-\tilde{x}_{\pi}\left(x_{b},x_{\Sigma}\right)\right)\right)\\&\;+xU_x\left(x,t\right)\left(\tilde{x}_{\pi}\left(x_{b},x_{\Sigma}\right)+\eta\left(\hat{x}_{\pi}-\tilde{x}_{\pi}\left(x_{b},x_{\Sigma}\right)\right)\right)^{tr}\left(x_{b}-r\mathds{1}\right)\\=&\;F\left(x,t;\tilde{x}_{\pi}\left(x_{b},x_{\Sigma}\right);x_{b},x_{\Sigma}\right)+\eta x^2U_{xx}\left(x,t\right)\left(\hat{x}_{\pi}-\tilde{x}_{\pi}\left(x_{b},x_{\Sigma}\right)\right)^{tr}x_{\Sigma}\tilde{x}_{\pi}\left(x_{b},x_{\Sigma}\right)\\&\;+\frac{1}{2}\eta^2 x^2U_{xx}\left(x,t\right)\left(\hat{x}_{\pi}-\tilde{x}_{\pi}\left(x_{b},x_{\Sigma}\right)\right)^{tr}x_{\Sigma}\left(\hat{x}_{\pi}-\tilde{x}_{\pi}\left(x_{b},x_{\Sigma}\right)\right)\\&\;+\eta xU_x\left(x,t\right)\left(\hat{x}_{\pi}-\tilde{x}_{\pi}\left(x_{b},x_{\Sigma}\right)\right)^{tr}\left(x_{b}-r\mathds{1}\right).
\end{align*}
Therefore, by letting $\eta\rightarrow 0$, for any $\hat{x}_{\pi}\in\Pi$,
\begin{equation*}
\left(\hat{x}_{\pi}-\tilde{x}_{\pi}\left(x_{b},x_{\Sigma}\right)\right)^{tr}\left(xU_{xx}\left(x,t\right)x_{\Sigma}\tilde{x}_{\pi}\left(x_{b},x_{\Sigma}\right)+U_x\left(x,t\right)\left(x_{b}-r\mathds{1}\right)\right)\leq 0,
\end{equation*}
which holds in particular for $\hat{x}_{\pi}=\tilde{x}_{\pi}\left(x'_{b},x'_{\Sigma}\right)$; that is,
\begin{equation*}
\left(\tilde{x}_{\pi}\left(x'_{b},x'_{\Sigma}\right)-\tilde{x}_{\pi}\left(x_{b},x_{\Sigma}\right)\right)^{tr}\left(xU_{xx}\left(x,t\right)x_{\Sigma}\tilde{x}_{\pi}\left(x_{b},x_{\Sigma}\right)+U_x\left(x,t\right)\left(x_{b}-r\mathds{1}\right)\right)\leq 0.
\end{equation*}
Similarly,
\begin{equation*}
\left(\tilde{x}_{\pi}\left(x_{b},x_{\Sigma}\right)-\tilde{x}_{\pi}\left(x'_{b},x'_{\Sigma}\right)\right)^{tr}\left(xU_{xx}\left(x,t\right)x'_{\Sigma}\tilde{x}_{\pi}\left(x'_{b},x'_{\Sigma}\right)+U_x\left(x,t\right)\left(x'_{b}-r\mathds{1}\right)\right)\leq 0.
\end{equation*}
Their sum yields
\begin{align*}
\left(\tilde{x}_{\pi}\left(x'_{b},x'_{\Sigma}\right)-\tilde{x}_{\pi}\left(x_{b},x_{\Sigma}\right)\right)^{tr}\big(&xU_{xx}\left(x,t\right)\left(x_{\Sigma}\tilde{x}_{\pi}\left(x_{b},x_{\Sigma}\right)-x'_{\Sigma}\tilde{x}_{\pi}\left(x'_{b},x'_{\Sigma}\right)\right)\\&+U_x\left(x,t\right)\left(x_{b}-x'_{b}\right)\big)\leq 0,
\end{align*}
which implies that, by telescoping,
\begin{equation}
\begin{aligned}
&\;\left(\tilde{x}_{\pi}\left(x'_{b},x'_{\Sigma}\right)-\tilde{x}_{\pi}\left(x_{b},x_{\Sigma}\right)\right)^{tr}x'_{\Sigma}\left(\tilde{x}_{\pi}\left(x'_{b},x'_{\Sigma}\right)-\tilde{x}_{\pi}\left(x_{b},x_{\Sigma}\right)\right)\\\leq&\;\left(\tilde{x}_{\pi}\left(x'_{b},x'_{\Sigma}\right)-\tilde{x}_{\pi}\left(x_{b},x_{\Sigma}\right)\right)^{tr}\left(-\frac{U_x\left(x,t\right)}{xU_{xx}\left(x,t\right)}\left(x'_{b}-x_{b}\right)-\left(x'_{\Sigma}-x_{\Sigma}\right)\tilde{x}_{\pi}\left(x_{b},x_{\Sigma}\right)\right).
\end{aligned}
\label{eq:telescoping}
\end{equation}
Since $x'_{\Sigma}\in\Sigma$ is a real symmetric positive-definite matrix, by the Cholesky decomposition, $x'_{\Sigma}=\left(x'_{\sigma}\right)^{tr}x'_{\sigma}$, where $x'_{\sigma}$ is a unique $d\times d$ upper triangular matrix with real and positive diagonal entries. Hence, by the Cauchy Schwarz inequality and \eqref{eq:telescoping},
\begin{align*}
&\;\vert\tilde{x}_{\pi}\left(x'_{b},x'_{\Sigma}\right)-\tilde{x}_{\pi}\left(x_{b},x_{\Sigma}\right)\vert^2\\=&\;\vert\left(x'_{\sigma}\right)^{-1}x'_{\sigma}\left(\tilde{x}_{\pi}\left(x'_{b},x'_{\Sigma}\right)-\tilde{x}_{\pi}\left(x_{b},x_{\Sigma}\right)\right)\vert^2\\\leq&\;\vert\left(x'_{\sigma}\right)^{-1}\vert^2\vert x'_{\sigma}\left(\tilde{x}_{\pi}\left(x'_{b},x'_{\Sigma}\right)-\tilde{x}_{\pi}\left(x_{b},x_{\Sigma}\right)\right)\vert^2\\=&\;\vert\left(x'_{\sigma}\right)^{-1}\vert^2\left(x'_{\sigma}\left(\tilde{x}_{\pi}\left(x'_{b},x'_{\Sigma}\right)-\tilde{x}_{\pi}\left(x_{b},x_{\Sigma}\right)\right)\right)^{tr}x'_{\sigma}\left(\tilde{x}_{\pi}\left(x'_{b},x'_{\Sigma}\right)-\tilde{x}_{\pi}\left(x_{b},x_{\Sigma}\right)\right)\\=&\;\vert\left(x'_{\sigma}\right)^{-1}\vert^2\left(\tilde{x}_{\pi}\left(x'_{b},x'_{\Sigma}\right)-\tilde{x}_{\pi}\left(x_{b},x_{\Sigma}\right)\right)^{tr}x'_{\Sigma}\left(\tilde{x}_{\pi}\left(x'_{b},x'_{\Sigma}\right)-\tilde{x}_{\pi}\left(x_{b},x_{\Sigma}\right)\right)\\\leq&\;\vert\left(x'_{\sigma}\right)^{-1}\vert^2\left(\tilde{x}_{\pi}\left(x'_{b},x'_{\Sigma}\right)-\tilde{x}_{\pi}\left(x_{b},x_{\Sigma}\right)\right)^{tr}\\&\;\left(-\frac{U_x\left(x,t\right)}{xU_{xx}\left(x,t\right)}\left(x'_{b}-x_{b}\right)-\left(x'_{\Sigma}-x_{\Sigma}\right)\tilde{x}_{\pi}\left(x_{b},x_{\Sigma}\right)\right)\\\leq&\;\vert\left(x'_{\sigma}\right)^{-1}\vert^2\vert\tilde{x}_{\pi}\left(x'_{b},x'_{\Sigma}\right)-\tilde{x}_{\pi}\left(x_{b},x_{\Sigma}\right)\vert\\&\;\times\left(\bigg\vert\frac{U_x\left(x,t\right)}{xU_{xx}\left(x,t\right)}\bigg\vert\vert x'_{b}-x_{b}\vert+\vert x'_{\Sigma}-x_{\Sigma}\vert\vert\tilde{x}_{\pi}\left(x_{b},x_{\Sigma}\right)\vert\right);
\end{align*}
that is,
\begin{equation}
\begin{aligned}
&\;\vert\tilde{x}_{\pi}\left(x'_{b},x'_{\Sigma}\right)-\tilde{x}_{\pi}\left(x_{b},x_{\Sigma}\right)\vert\\\leq&\;\vert\left(x'_{\sigma}\right)^{-1}\vert^2\left(\bigg\vert\frac{U_x\left(x,t\right)}{xU_{xx}\left(x,t\right)}\bigg\vert\vert x'_{b}-x_{b}\vert+\vert x'_{\Sigma}-x_{\Sigma}\vert\vert\tilde{x}_{\pi}\left(x_{b},x_{\Sigma}\right)\vert\right).
\end{aligned}
\label{eq:prior_estimate}
\end{equation}
Let $\left(x_{b,0},x_{\Sigma,0}\right)\in\mathbb{B}\times\Sigma$, and let a sequence $\left\{\left(x_{b,n},x_{\Sigma,n}\right)\right\}_{n=1}^{\infty}\in\mathbb{B}\times\Sigma$ which converges to $\left(x_{b,0},x_{\Sigma,0}\right)$. By \eqref{eq:prior_estimate},
\begin{align*}
&\;\vert\tilde{x}_{\pi}\left(x_{b,n},x_{\Sigma,n}\right)-\tilde{x}_{\pi}\left(x_{b,0},x_{\Sigma,0}\right)\vert\\\leq&\;\vert x_{\sigma,n}^{-1}\vert^2\left(\bigg\vert\frac{U_x\left(x,t\right)}{xU_{xx}\left(x,t\right)}\bigg\vert\vert x_{b,n}-x_{b,0}\vert+\vert x_{\Sigma,n}-x_{\Sigma,0}\vert\vert\tilde{x}_{\pi}\left(x_{b,0},x_{\Sigma,0}\right)\vert\right),
\end{align*}
where $x_{\sigma,n}$ is a unique $d\times d$ upper triangular matrix with real and positive diagonal entries, such that $x_{\sigma,n}^{tr}x_{\sigma,n}=x_{\Sigma,n}$, for $n=1,2,\dots$, by the Cholesky decomposition.
Due to the compactness, and the consequent boundedness, of $\Sigma$, and by the bounded inverse theorem as well as the equivalence of norms on finite-dimensional vector spaces, there exists a positive constant $M>0$ such that $\vert x_{\sigma,n}^{-1}\vert\leq M$ uniformly for all $n=1,2,\dots$. Therefore, the sequence $\left\{\tilde{x}_{\pi}\left(x_{b,n},x_{\Sigma,n}\right)\right\}_{n=1}^{\infty}$ also converges to $\tilde{x}_{\pi}\left(x_{b,0},x_{\Sigma,0}\right)$. All these together prove that any maximizer $\tilde{x}_{\pi}\left(x_{b},x_{\Sigma}\right)\in\Pi$ of \eqref{eq:F} is continuous in $\left(x_{b},x_{\Sigma}\right)\in\mathbb{B}\times\Sigma$.

By such continuity,
the function, for any $\left(x_{b},x_{\Sigma}\right)\in\mathbb{B}\times\Sigma$,
\begin{align*}
&\;\tilde{F}\left(x,t;x_{b},x_{\Sigma}\right)=F\left(x,t;\tilde{x}_{\pi}\left(x_{b},x_{\Sigma}\right);x_{b},x_{\Sigma}\right)\\=&\;\frac{1}{2}x^2U_{xx}\left(x,t\right)\left(\tilde{x}_{\pi}\left(x_{b},x_{\Sigma}\right)\right)^{tr}x_{\Sigma}\tilde{x}_{\pi}\left(x_{b},x_{\Sigma}\right)+xU_x\left(x,t\right)\left(\tilde{x}_{\pi}\left(x_{b},x_{\Sigma}\right)\right)^{tr}\left(x_{b}-r\mathds{1}\right)
\end{align*}
is also continuous in $\left(x_{b},x_{\Sigma}\right)\in\mathbb{B}\times\Sigma$. As the sets $\mathbb{B}$ and $\Sigma$ are compact,
\begin{equation}\label{minimizer}
\left(x^*_{b},x^*_{\Sigma}\right)=\argmin_{\left(x_{b},x_{\Sigma}\right)\in\mathbb{B}\times\Sigma}\tilde{F}\left(x,t;x_{b},x_{\Sigma}\right)
\end{equation}
exists by the extreme value theorem. Also, with $\tilde{x}_{\pi}(\cdot,\cdot)$ given in (\ref{eq:tilde_pi}), define
\begin{equation}\label{maximizer}
x^*_{\pi}=\tilde{x}_{\pi}\left(x^*_{b},x^*_{\Sigma}\right)\in\Pi.
\end{equation}

\underline{Step 3}. Then, $\left(x^*_{\pi};x^*_{b},x^*_{\Sigma}\right)\in\Pi\times\mathbb{B}\times\Sigma$, which depends on $\left(x,t\right)\in \mathbb{R}_{++}\times\left[0,\infty\right)$, is a saddle-point of $F\left(x,t;\cdot;\cdot,\cdot\right)$ in \eqref{eq:F}. By the definition of $x^*_{\pi}=\tilde{x}_{\pi}\left(x^*_{b},x^*_{\Sigma}\right)\in\Pi$ maximizing $F\left(x,t;\cdot;x^*_{b},x^*_{\Sigma}\right)$, for any $x_{\pi}\in\Pi$,
\begin{equation*}
F\left(x,t;x_{\pi};x^*_{b},x^*_{\Sigma}\right)\leq F\left(x,t;\tilde{x}_{\pi}\left(x^*_{b},x^*_{\Sigma}\right);x^*_{b},x^*_{\Sigma}\right)=F\left(x,t;x^*_{\pi};x^*_{b},x^*_{\Sigma}\right).
\end{equation*}
Let $\left(x_{b},x_{\Sigma}\right)\in\mathbb{B}\times\Sigma$. By the convexity of $\mathbb{B}$ and $\Sigma$, for any $\gamma\in\left(0,1\right]$, $\left(x_{b,\gamma},x_{\Sigma,\gamma}\right)=\gamma\left(x_{b},x_{\Sigma}\right)+\left(1-\gamma\right)\left(x^*_{b},x^*_{\Sigma}\right)\in\mathbb{B}\times\Sigma$, and thus
\begin{align*}
&\;F\left(x,t;x^*_{\pi};x^*_{b},x^*_{\Sigma}\right)=F\left(x,t;\tilde{x}_{\pi}\left(x^*_{b},x^*_{\Sigma}\right);x^*_{b},x^*_{\Sigma}\right)=\tilde{F}\left(x,t;x^*_{b},x^*_{\Sigma}\right)\\\leq&\;\tilde{F}\left(x,t;x_{b,\gamma},x_{\Sigma,\gamma}\right)=F\left(x,t;\tilde{x}_{\pi}\left(x_{b,\gamma},x_{\Sigma,\gamma}\right);x_{b,\gamma},x_{\Sigma,\gamma}\right)\\=&\;\gamma F\left(x,t;\tilde{x}_{\pi}\left(x_{b,\gamma},x_{\Sigma,\gamma}\right);x_{b},x_{\Sigma}\right)+\left(1-\gamma\right)F\left(x,t;\tilde{x}_{\pi}\left(x_{b,\gamma},x_{\Sigma,\gamma}\right);x^*_{b},x^*_{\Sigma}\right)\\\leq&\;\gamma F\left(x,t;\tilde{x}_{\pi}\left(x_{b,\gamma},x_{\Sigma,\gamma}\right);x_{b},x_{\Sigma}\right)+\left(1-\gamma\right)F\left(x,t;\tilde{x}_{\pi}\left(x^*_{b},x^*_{\Sigma}\right);x^*_{b},x^*_{\Sigma}\right)\\=&\;\gamma F\left(x,t;\tilde{x}_{\pi}\left(x_{b,\gamma},x_{\Sigma,\gamma}\right);x_{b},x_{\Sigma}\right)+\left(1-\gamma\right)F\left(x,t;x^*_{\pi};x^*_{b},x^*_{\Sigma}\right),
\end{align*}
where the first and fifth equalities are due to the definition of $x^*_{\pi}=\tilde{x}_{\pi}\left(x^*_{b},x^*_{\Sigma}\right)$, the second and third equalities are by the definition of $\tilde{F}\left(x,t;\cdot,\cdot\right)$, the fourth equality is due to the linearity of $F\left(x,t;x_{\pi};\cdot,\cdot\right)$ for any $x_{\pi}\in\Pi$, the first inequality is by the definition of $\left(x^*_{b},x^*_{\Sigma}\right)=\argmin_{\left(x_{b},x_{\Sigma}\right)\in\mathbb{B}\times\Sigma}\tilde{F}\left(x,t;x_{b},x_{\Sigma}\right)$, and the second inequality is due to the definition of $x^*_{\pi}=\tilde{x}_{\pi}\left(x^*_{b},x^*_{\Sigma}\right)\in\Pi$ maximizing $F\left(x,t;\cdot;x^*_{b},x^*_{\Sigma}\right)$ for any $x_{\pi}\in\Pi$. Therefore, for any $\gamma\in\left(0,1\right]$,
\begin{equation*}
F\left(x,t;x^*_{\pi};x^*_{b},x^*_{\Sigma}\right)\leq F\left(x,t;\tilde{x}_{\pi}\left(x_{b,\gamma},x_{\Sigma,\gamma}\right);x_{b},x_{\Sigma}\right).
\end{equation*}
Finally, when $\gamma\rightarrow 0$, $\left(x_{b,\gamma},x_{\Sigma,\gamma}\right)$ converges to $\left(x^*_{b},x^*_{\Sigma}\right)$, and thus, by the continuity of $\tilde{x}_{\pi}\left(\cdot,\cdot\right)$ in $\mathbb{B}\times\Sigma$, and of $F\left(x,t;\cdot;x_b,x_{\Sigma}\right)$ for any $\left(x_b,x_{\Sigma}\right)\in\mathbb{B}\times\Sigma$, as well as the definition of $x^*_{\pi}=\tilde{x}_{\pi}\left(x^*_{b},x^*_{\Sigma}\right)$,
\begin{align*}
F\left(x,t;x^*_{\pi};x^*_{b},x^*_{\Sigma}\right)\leq&\; \lim_{\gamma\rightarrow 0}F\left(x,t;\tilde{x}_{\pi}\left(x_{b,\gamma},x_{\Sigma,\gamma}\right);x_{b},x_{\Sigma}\right)\\=&\;F\left(x,t;x^*_{\pi};x_{b},x_{\Sigma}\right).
\end{align*}

\section{Proof of Theorem \ref{prop:SPDE}}
\underline{Step 1}. Let $t\geq 0$, $\xi\in\mathcal{L}\left(\mathcal{F}_t;\mathbb{R}_{++}\right)$, and $T\geq t$. For any $\left(\pi,c\right)\in\mathcal{A}$, $\left(b,\sigma\right)\in\mathcal{B}$, and $s\geq t$, define
\begin{equation*}
R_s^{\xi,t;\pi,c;b,\sigma}=U\left(X_s^{\xi,t;\pi,c;b,\sigma},s\right)+\int_{t}^{s}U^c\left(c_vX_v^{\xi,t;\pi,c;b,\sigma},v\right)dv;
\end{equation*}
in particular, $R_t^{\xi,t;\pi,c;b,\sigma}=U\left(\xi,t\right)$. By It{\^o}'s formula, $R^{\xi,t;\pi,c;b,\sigma}$ solves, for any $s\geq t$,
\begin{equation}
\begin{aligned}
dR_s=&\left(\frac{1}{2}X_s^2U_{xx}\left(X_s,s\right)\pi_s^{tr}\sigma_s\sigma_s^{tr}\pi_s+X_sU_x\left(X_s,s\right)\pi_s^{tr}\left(b_s-r\mathds{1}\right)\right.\\&\;\;\left.-F\left(X_s,s;x^*_{\pi};x^*_{b},x^*_{\Sigma}\right)+\left(U^c\left(c_sX_s,s\right)-c_sX_sU_x\left(X_s,s\right)\right)\right.\\&\;\;\left.-\sup_{x_c\in\mathbb{R}_+}\left(U^c\left(x_cX_s,s\right)-x_cX_sU_x\left(X_s,s\right)\right)\right)ds+X_sU_x\left(X_s,s\right)\pi_s^{tr}\sigma_sdW_s.
\end{aligned}
\label{eq:R_new}
\end{equation}

Define, for any $t\geq 0$, $b^*_t=x^*_{b}\left(X_t,t\right)$. Also, for any $t\geq 0$, since $x^*_{\Sigma}\left(X_t,t\right)$ is a real symmetric positive-definite matrix, by the Cholesky decomposition, $x^*_{\Sigma}\left(X_t,t\right)=x^*_{\sigma}\left(X_t,t\right)\left(x^*_{\sigma}\left(X_t,t\right)\right)^{tr}$, where $x^*_{\sigma}\left(X_t,t\right)$ is a unique $d\times d$ lower triangular matrix with real and positive diagonal entries; define $\sigma^*_t=x^*_{\sigma}\left(X_t,t\right)$. Note that $\left(b^*,\sigma^*\right)\in\mathcal{B}$.

\underline{Step 2}. First, note that for any $\left(\pi,c\right)\in\mathcal{A}$ and $s\geq t$,
\begin{align*}
&\;\frac{1}{2}X_s^2U_{xx}\left(X_s,s\right)\pi_s^{tr}\sigma^*_s\left(\sigma^*_s\right)^{tr}\pi_s+X_sU_x\left(X_s,s\right)\pi_s^{tr}\left(b^*_s-r\mathds{1}\right)\\=&\;\frac{1}{2}X_s^2U_{xx}\left(X_s,s\right)\pi_s^{tr}x^*_{\Sigma}\left(X_s,s\right)\pi_s+X_sU_x\left(X_s,s\right)\pi_s^{tr}\left(x^*_{b}\left(X_s,s\right)-r\mathds{1}\right)\\=&\;F\left(X_s,s;\pi_s;x^*_{b},x^*_{\Sigma}\right)\\\leq&\; F\left(X_s,s;x^*_{\pi};x^*_{b},x^*_{\Sigma}\right),
\end{align*}
due to \eqref{eq:saddle_inequality}; and,
\begin{equation*}
U^c\left(c_sX_s,s\right)-c_sX_sU_x\left(X_s,s\right)\leq\sup_{x_c\in\mathbb{R}_+}\left(U^c\left(x_cX_s,s\right)-x_cX_sU_x\left(X_s,s\right)\right).
\end{equation*}
Hence, by \eqref{eq:R_new}, for any $\left(\pi,c\right)\in\mathcal{A}$, $R^{\xi,t;\pi,c;b^*,\sigma^*}$ is an $\mathbb{F}$-local supermartingale. Note that $R^{\xi,t;\pi,c;b^*,\sigma^*}$ is bounded from below; indeed, for any $s\geq t$,
\begin{align*}
R_s^{\xi,t;\pi,c;b^*,\sigma^*}\geq&\; M_1+\int_{t}^{s}U^c\left(0,v\right)dv\\=&\;M_1+\int_{t}^{s}\max\left\{U^c\left(0,v\right),0\right\}dv+\int_{t}^{s}\min\left\{U^c\left(0,v\right),0\right\}dv\\\geq&\;M_1+0+\int_{0}^{\infty}\min\left\{U^c\left(0,v\right),0\right\}dv\\\geq&\;M_1+M_2.
\end{align*}
Therefore, it is a proper $\mathbb{F}$-supermartingale. In particular, for any $\left(\pi,c\right)\in\mathcal{A}$, $R_t^{\xi,t;\pi,c;b^*,\sigma^*}\geq\mathbb{E}\left[R_T^{\xi,t;\pi,c;b^*,\sigma^*}\vert\mathcal{F}_t\right]$; that is, for any $\left(\pi,c\right)\in\mathcal{A}$,
\begin{equation}
\begin{aligned}
&\;U\left(\xi,t\right)\\\geq&\;\mathbb{E}\left[U\left(X_T^{\xi,t;\pi,c;b^*,\sigma^*},T\right)+\int_{t}^{T}U^c\left(c_sX_s^{\xi,t;\pi,c;b^*,\sigma^*},s\right)ds\vert\mathcal{F}_t\right]\\\geq&\;\essinf_{\left(b,\sigma\right)\in\mathcal{B}}\mathbb{E}\left[U\left(X_T^{\xi,t;\pi,c;b,\sigma},T\right)+\int_{t}^{T}U^c\left(c_sX_s^{\xi,t;\pi,c;b,\sigma},s\right)ds\vert\mathcal{F}_t\right],
\end{aligned}
\label{eq:combine_1}
\end{equation}
which holds particularly for $\left(\pi^*,c^*\right)\in\mathcal{A}$ given in \eqref{eq:optimal_strategies}, and which further implies that
\begin{equation}
\begin{aligned}
&\;U\left(\xi,t\right)\\\geq&\;\esssup_{\left(\pi,c\right)\in\mathcal{A}}\mathbb{E}\left[U\left(X_T^{\xi,t;\pi,c;b^*,\sigma^*},T\right)+\int_{t}^{T}U^c\left(c_sX_s^{\xi,t;\pi,c;b^*,\sigma^*},s\right)ds\vert\mathcal{F}_t\right]\\\geq&\;\essinf_{\left(b,\sigma\right)\in\mathcal{B}}\esssup_{\left(\pi,c\right)\in\mathcal{A}}\mathbb{E}\left[U\left(X_T^{\xi,t;\pi,c;b,\sigma},T\right)+\int_{t}^{T}U^c\left(c_sX_s^{\xi,t;\pi,c;b,\sigma},s\right)ds\vert\mathcal{F}_t\right]\\\geq&\;\esssup_{\left(\pi,c\right)\in\mathcal{A}}\essinf_{\left(b,\sigma\right)\in\mathcal{B}}\mathbb{E}\left[U\left(X_T^{\xi,t;\pi,c;b,\sigma},T\right)+\int_{t}^{T}U^c\left(c_sX_s^{\xi,t;\pi,c;b,\sigma},s\right)ds\vert\mathcal{F}_t\right],
\end{aligned}
\label{eq:combine_2}
\end{equation}
where the last inequality is due to the max-min inequality.

\underline{Step 3}. Next, we prove the other side of the above inequality. For any $\left(b,\sigma\right)\in\mathcal{B}$ and $s\geq t$,
\begin{align*}
&\;\frac{1}{2}X_s^2U_{xx}\left(X_s,s\right)\left(\pi^*_s\right)^{tr}\sigma_s\sigma_s^{tr}\pi^*_s+X_sU_x\left(X_s,s\right)\left(\pi^*_s\right)^{tr}\left(b_s-r\mathds{1}\right)\\=&\;\frac{1}{2}X_s^2U_{xx}\left(X_s,s\right)\left(x^*_{\pi}\left(X_s,s\right)\right)^{tr}\sigma_s\sigma_s^{tr}x^*_{\pi}\left(X_s,s\right)\\&+X_sU_x\left(X_s,s\right)\left(x^*_{\pi}\left(X_s,s\right)\right)^{tr}\left(b_s-r\mathds{1}\right)\\=&\;F\left(X_s,s;x^*_{\pi};b_s,\sigma_s\sigma_s^{tr}\right)\\\geq&\;F\left(X_s,s;x^*_{\pi};x^*_{b},x^*_{\Sigma}\right),
\end{align*}
due to \eqref{eq:saddle_inequality}; and,
\begin{equation*}
U^c\left(c^*_sX_s,s\right)-c^*_sX_sU_x\left(X_s,s\right)=\sup_{x_c\in\mathbb{R}_+}\left(U^c\left(x_cX_s,s\right)-x_cX_sU_x\left(X_s,s\right)\right).
\end{equation*}
Hence, by \eqref{eq:R_new}, for any $\left(b,\sigma\right)\in\mathcal{B}$, $R^{\xi,t;\pi^*,c^*;b,\sigma}$ is an $\mathbb{F}$-local submartingale; that is, there exists a sequence $\left\{\tau_n\right\}_{n=1}^{\infty}$ of $\mathbb{F}$-stopping times such that, $\tau_n\geq t$ for all $n=1,2,\dots$, $\tau_n<\tau_{n+1}$, $\mathbb{P}$-a.s., for all $n=1,2,\dots$, $\tau_n\rightarrow\infty$ as $n\rightarrow\infty$, $\mathbb{P}$-a.s., and $R_{\tau_n\wedge\cdot}^{\xi,t;\pi^*,c^*;b,\sigma}$ is an $\mathbb{F}$-submartingale. In particular, for any $\left(b,\sigma\right)\in\mathcal{B}$, $n=1,2,\dots$, and $s\geq t$, $R_t^{\xi,t;\pi^*,c^*;b,\sigma}\leq\mathbb{E}\left[R_{\tau_n\wedge s}^{\xi,t;\pi^*,c^*;b,\sigma}\vert\mathcal{F}_t\right]$; that is, for any $\left(b,\sigma\right)\in\mathcal{B}$, $n=1,2,\dots$, and $s\geq t$,
\begin{equation}
\begin{aligned}
U\left(\xi,t\right)\leq&\;\mathbb{E}\left[U\left(X_{\tau_n\wedge s}^{\xi,t;\pi^*,c^*;b,\sigma},\tau_n\wedge s\right)\right.\\&\quad\quad\left.+\int_{t}^{\tau_n\wedge s}U^c\left(c^*_vX_v^{\xi,t;\pi^*,c^*;b,\sigma},v\right)dv\vert\mathcal{F}_t\right].
\end{aligned}
\label{eq:MCT_BCT_eq}
\end{equation}
Recall that, for any $\left(b,\sigma\right)\in\mathcal{B}$ and $t\geq 0$, $\left\{U\left(X^{\xi;\pi^*,c^*;b,\sigma}_{\tau},\tau\right)\right\}_{\tau\in\mathcal{T}\left[0,t\right]}$ is uniformly integrable, where $\mathcal{T}\left[0,t\right]$ is the set of all $\mathbb{F}$-stopping time $\tau\in\left[0,t\right]$. Moreover, for any $\left(b,\sigma\right)\in\mathcal{B}$, $n=1,2,\dots$, and $s\geq t$,
\begin{align*}
0\geq&\;\int_{t}^{\tau_n\wedge s}\min\left\{U^c\left(c^*_vX_v^{\xi,t;\pi^*,c^*;b,\sigma},v\right),0\right\}dv\\\geq&\;\int_{0}^{\infty}\min\left\{U^c\left(0,v\right),0\right\}dv\geq M_2.
\end{align*}
Therefore, by \eqref{eq:MCT_BCT_eq}, the bounded convergence theorem, and the monotone convergence theorem, for any $\left(b,\sigma\right)\in\mathcal{B}$ and $s\geq t$,
\begin{align*}
U\left(\xi,t\right)\leq&\;\lim_{n\rightarrow\infty}\mathbb{E}\left[U\left(X_{\tau_n\wedge s}^{\xi,t;\pi^*,c^*;b,\sigma},\tau_n\wedge s\right)\right.\\&\quad\quad\quad\quad\left.+\int_{t}^{\tau_n\wedge s}\min\left\{U^c\left(c^*_vX_v^{\xi,t;\pi^*,c^*;b,\sigma},v\right),0\right\}dv\right.\\&\quad\quad\quad\quad\left.+\int_{t}^{\tau_n\wedge s}\max\left\{U^c\left(c^*_vX_v^{\xi,t;\pi^*,c^*;b,\sigma},v\right),0\right\}dv\vert\mathcal{F}_t\right]\\\leq&\;\mathbb{E}\left[U\left(X_s^{\xi,t;\pi^*,c^*;b,\sigma},s\right)+\int_{t}^{s}\min\left\{U^c\left(c^*_vX_v^{\xi,t;\pi^*,c^*;b,\sigma},v\right),0\right\}dv\right.\\&\quad\quad\left.+\int_{t}^{s}\max\left\{U^c\left(c^*_vX_v^{\xi,t;\pi^*,c^*;b,\sigma},v\right),0\right\}dv\vert\mathcal{F}_t\right]\\=&\;\mathbb{E}\left[U\left(X_s^{\xi,t;\pi^*,c^*;b,\sigma},s\right)+\int_{t}^{s}U^c\left(c^*_vX_v^{\xi,t;\pi^*,c^*;b,\sigma},v\right)dv\vert\mathcal{F}_t\right].
\end{align*}
In particular, for any $\left(b,\sigma\right)\in\mathcal{B}$,
\begin{equation}
\begin{aligned}
&\;U\left(\xi,t\right)\\\leq&\;\mathbb{E}\left[U\left(X_T^{\xi,t;\pi^*,c^*;b,\sigma},T\right)+\int_{t}^{T}U^c\left(c^*_sX_s^{\xi,t;\pi^*,c^*;b,\sigma},s\right)ds\vert\mathcal{F}_t\right]\\\leq&\;\esssup_{\left(\pi,c\right)\in\mathcal{A}}\mathbb{E}\left[U\left(X_T^{\xi,t;\pi,c;b,\sigma},T\right)+\int_{t}^{T}U^c\left(c_sX_s^{\xi,t;\pi,c;b,\sigma},s\right)ds\vert\mathcal{F}_t\right],
\end{aligned}
\label{eq:combine_3}
\end{equation}
which holds particularly for $\left(b^*,\sigma^*\right)\in\mathcal{B}$, and which further implies that
\begin{equation}
\begin{aligned}
&\;U\left(\xi,t\right)\\\leq&\;\essinf_{\left(b,\sigma\right)\in\mathcal{B}}\mathbb{E}\left[U\left(X_T^{\xi,t;\pi^*,c^*;b,\sigma},T\right)+\int_{t}^{T}U^c\left(c^*_sX_s^{\xi,t;\pi^*,c^*;b,\sigma},s\right)ds\vert\mathcal{F}_t\right]\\\leq&\;\esssup_{\left(\pi,c\right)\in\mathcal{A}}\essinf_{\left(b,\sigma\right)\in\mathcal{B}}\mathbb{E}\left[U\left(X_T^{\xi,t;\pi,c;b,\sigma},T\right)+\int_{t}^{T}U^c\left(c_sX_s^{\xi,t;\pi,c;b,\sigma},s\right)ds\vert\mathcal{F}_t\right]\\\leq&\;\essinf_{\left(b,\sigma\right)\in\mathcal{B}}\esssup_{\left(\pi,c\right)\in\mathcal{A}}\mathbb{E}\left[U\left(X_T^{\xi,t;\pi,c;b,\sigma},T\right)+\int_{t}^{T}U^c\left(c_sX_s^{\xi,t;\pi,c;b,\sigma},s\right)ds\vert\mathcal{F}_t\right],
\end{aligned}
\label{eq:combine_4}
\end{equation}
where the last inequality is due to the max-min inequality.

\underline{Step 4}. Finally, by \eqref{eq:combine_1} with $\left(\pi^*,c^*\right)\in\mathcal{A}$, \eqref{eq:combine_2}, \eqref{eq:combine_3} with $\left(b^*,\sigma^*\right)\in\mathcal{B}$, and \eqref{eq:combine_4},
\begin{align*}
&\;U\left(\xi,t\right)\\=&\;\esssup_{\left(\pi,c\right)\in\mathcal{A}}\essinf_{\left(b,\sigma\right)\in\mathcal{B}}\mathbb{E}\left[U\left(X_T^{\xi,t;\pi,c;b,\sigma},T\right)+\int_{t}^{T}U^c\left(c_sX_s^{\xi,t;\pi,c;b,\sigma},s\right)ds\vert\mathcal{F}_t\right]\\=&\;\essinf_{\left(b,\sigma\right)\in\mathcal{B}}\mathbb{E}\left[U\left(X_T^{\xi,t;\pi^*,c^*;b,\sigma},T\right)+\int_{t}^{T}U^c\left(c^*_sX_s^{\xi,t;\pi^*,c^*;b,\sigma},s\right)ds\vert\mathcal{F}_t\right]\\=&\;\mathbb{E}\left[U\left(X_T^{\xi,t;\pi^*,c^*;b^*,\sigma^*},T\right)+\int_{t}^{T}U^c\left(c^*_sX_s^{\xi,t;\pi^*,c^*;b^*,\sigma^*},s\right)ds\vert\mathcal{F}_t\right]\\=&\;\esssup_{\left(\pi,c\right)\in\mathcal{A}}\mathbb{E}\left[U\left(X_T^{\xi,t;\pi,c;b^*,\sigma^*},T\right)+\int_{t}^{T}U^c\left(c_sX_s^{\xi,t;\pi,c;b^*,\sigma^*},s\right)ds\vert\mathcal{F}_t\right]\\=&\;\essinf_{\left(b,\sigma\right)\in\mathcal{B}}\esssup_{\left(\pi,c\right)\in\mathcal{A}}\mathbb{E}\left[U\left(X_T^{\xi,t;\pi,c;b,\sigma},T\right)+\int_{t}^{T}U^c\left(c_sX_s^{\xi,t;\pi,c;b,\sigma},s\right)ds\vert\mathcal{F}_t\right].
\end{align*}

\section{Proof of Proposition \ref{theorem:new_theorem_1}}

It is clear that $U\left(\cdot,\cdot\right)$ and $U^c\left(\cdot,\cdot\right)$ given by \eqref{eq:forward_utility_drift_vol_uncertainty} are non-decreasing and concave, in $x\in\mathbb{R}_{++}$ and $x_C\in\mathbb{R}_+$ respectively, and they are in $C^{2,1}$.

With $U\left(\cdot,\cdot\right)$ given by \eqref{eq:forward_utility_drift_vol_uncertainty}, and with a generic deterministic function $Y_t$, $t\geq 0$, for any $\left(x_{\pi};x_{b},x_{\Sigma}\right)\in\Pi\times\mathbb{B}\times\Sigma$,
\begin{equation*}
G\left(x_{\pi};x_{b},x_{\Sigma}\right)=U\left(x,t\right)^{-1}F\left(x,t;x_{\pi};x_{b},x_{\Sigma}\right),
\end{equation*}
for all $x\in\mathbb{R}_{++}$ and $t\geq 0$, where the deterministic function $F\left(x,t;\cdot;\cdot,\cdot\right)$ is given in \eqref{eq:F}. By Lemma \ref{lemma:saddle_point}, for any $x\in\mathbb{R}_{++}$ and $t\geq 0$, the deterministic function $F\left(x,t;\cdot;\cdot,\cdot\right)$ admits a saddle-point $\left(x^*_{\pi};x^*_{b},x^*_{\Sigma}\right)\in\Pi\times\mathbb{B}\times\Sigma$; hence, so does the deterministic function $G\left(\cdot;\cdot,\cdot\right)$ with the same saddle-point. Since the deterministic function $G\left(\cdot;\cdot,\cdot\right)$ is independent of $\left(x,t\right)\in \mathbb{R}_{++}\times\left[0,\infty\right)$, its saddle-point and saddle-value are also independent of $\left(x,t\right)$. Since $0\in\Pi$, $G=G\left(x^*_{\pi};x^*_{b},x^*_{\Sigma}\right)\geq G\left(0;x^*_{b},x^*_{\Sigma}\right)=0$.

With the condition \eqref{eq:condition_1}, the ODE \eqref{eq:new_ODE} can be uniquely solved. Indeed, by an exponential transformation that, for any $t\geq 0$, $\bar{Y}_t=e^{\frac{Y_t}{1-\delta}}$, the solution $Y_t$, $t\geq 0$, of the ODE \eqref{eq:new_ODE} is uniquely given by, for any $t\geq 0$,
\begin{equation}
Y_t=-\left(G+\delta r\right)t+\left(1-\delta\right)\ln\left(e^{\frac{Y_0}{1-\delta}}-\int_{0}^{t}e^{\frac{G+\delta r}{1-\delta}s}\lambda_s^{\frac{1}{1-\delta}}ds\right).
\label{eq:solution_Y_1}
\end{equation}

It remains to show that all conditions in Theorem \ref{prop:SPDE} are satisfied. First, for any $x\in\mathbb{R}_{++}$ and $t\geq 0$, $U\left(x,t\right)\geq 0$, and $\int_{0}^{\infty}\min\left\{U^c\left(0,s\right),0\right\}ds=0$, for $U\left(\cdot,\cdot\right)$ and $U^c\left(\cdot,\cdot\right)$ given in \eqref{eq:forward_utility_drift_vol_uncertainty}.

With $U\left(\cdot,\cdot\right)$ and $U^c\left(\cdot,\cdot\right)$ given by \eqref{eq:forward_utility_drift_vol_uncertainty}, $x^*_{\pi}\left(X_t,t\right)=x^*_{\pi}\in\Pi$ in \eqref{eq:optimal_strategies}, which is a constant and independent of $X_t\in\mathbb{R}_{++}$ and $t\geq 0$; since, for any $x\in\mathbb{R}_{++}$ and $t\geq 0$, $\lim_{x_c\rightarrow 0+}U^c_{x_C}\left(x_cx,t\right)=+\infty$ and $\lim_{x_c\rightarrow\infty}U^c_{x_C}\left(x_cx,t\right)=0$, $x^*_{c}\left(X_t,t\right)=x^*_{c}\left(t\right)=\lambda_t^{\frac{1}{1-\delta}}e^{-\frac{Y_t}{1-\delta}}\in\mathbb{R}_+$ in \eqref{eq:optimal_strategies}, which is a deterministic function in time $t\geq 0$ and independent of $X_t\in\mathbb{R}_{++}$, solves the equation $\tilde{x}_c^{\delta-1}\lambda_t=e^{Y_t}$ for $\tilde{x}_c\in\mathbb{R}_{++}$. Therefore, $\left(\pi^*_t,c^*_t\right)$, $t\geq 0$, in \eqref{eq:optimal_strategies} are given by \eqref{eq:optimal_robust_strategy} in this case. Since the solution $Y$, as shown in \eqref{eq:solution_Y_1}, of the ODE (\ref{eq:new_ODE}) is deterministic and continuous, for any $t\geq 0$, $Y$ is bounded on $\left[0,t\right]$, and so is $c^*$. This fact, together with $\pi^*$ being a constant, show that, for any $t\geq 0$, $\int_{0}^{t}\left(\vert \pi^*_s\vert^2+\vert c^*_s\vert\right)ds<\infty$, $\mathbb{P}$-a.s. Therefore, $\left(\pi^*,c^*\right)\in\mathcal{A}$.

Let $\left(b,\sigma\right)\in\mathcal{B}$ and $t\geq 0$. Since $\pi^*$ and $c^*$ are independent of the wealth state variable, by \eqref{wealth}, for any $s\in\left[0,t\right]$,
\begin{equation*}
\left(X_s^{\xi;\pi^*,c^*;b,\sigma}\right)^{\delta}=\xi^{\delta}e^{\int_{0}^{s}\left(G\left(\pi_v^{*};b_v,\sigma_v\sigma_v^{tr}\right)-\delta c^*_v+\delta r\right)dv}\mathcal{E}\left(\int_{0}^{\cdot}\delta \left(\pi_v^{*}\right)^{tr}\sigma_vdW_v\right)_s.
\end{equation*}
Since $\pi^*$ is a constant, and $\sigma_s\sigma_s^{tr}$, $s\in\left[0,t\right]$, is uniformly bounded by the compactness of $\Sigma$, the Dol{\'e}ans-Dade exponential $\mathbb{F}$-martingale $\mathcal{E}\left(\int_{0}^{\cdot}\delta \left(\pi_v^{*}\right)^{tr}\sigma_vdW_v\right)_s$, $s\in\left[0,t\right]$, is uniformly integrable. For any $s\in\left[0,t\right]$,
\begin{equation*}
G\left(\pi_s^{*};b_s,\sigma_s\sigma_s^{tr}\right)=\frac{1}{2}\delta\left(\delta-1\right)\left(\pi_s^{*}\right)^{tr}\sigma_s\sigma_s^{tr}\pi_s^{*}+\delta \left(\pi_s^{*}\right)^{tr}\left(b_s-r\mathds{1}\right),
\end{equation*}
is uniformly bounded on $\left[0,t\right]$, due to the continuity of $G\left(\pi_s^{*};\cdot,\cdot\right)$ as well as the compactness of $\mathbb{B}$ and $\Sigma$. Recall that $c^*$ is bounded on $\left[0,t\right]$. Therefore, for any $\left(b,\sigma\right)\in\mathcal{B}$ and $t\geq 0$, $\left\{\left(X^{\xi;\pi^*,c^*;b,\sigma}_{\tau}\right)^{\delta}\right\}_{\tau\in\mathcal{T}[0,t]}$ is uniformly integrable; together with the boundedness of $Y$ on $\left[0,t\right]$, $\left\{U\left(X^{\xi;\pi^*,c^*;b,\sigma}_{\tau},\tau\right)\right\}_{\tau\in\mathcal{T}\left[0,t\right]}$ is uniformly integrable.

Finally, $U\left(\cdot,\cdot\right)$ given in \eqref{eq:forward_utility_drift_vol_uncertainty} satisfies the PDE \eqref{eq:PDE}, with \eqref{eq:equality}. Indeed, for any $x\in\mathbb{R}_{++}$ and $t\geq 0$,
\begin{equation*}
U_t\left(x,t\right)=U\left(x,t\right)Y'_t=-U\left(x,t\right)\left(G+(1-\delta)\lambda_t^{\frac{1}{1-\delta}}e^{-\frac{Y_t}{1-\delta}}+\delta r\right);
\end{equation*}
\vspace{-0.5cm}
\begin{align*}
f\left(x,t\right)=&-U\left(x,t\right)G\left(x^*_{\pi};x^*_{b},x^*_{\Sigma}\right)-\left(U^c\left(x^*_c\left(t\right)x,t\right)-x^*_c\left(t\right)xU_x\left(x,t\right)\right)\\&-rxU_x\left(x,t\right)\\=&-U\left(x,t\right)G-U\left(x,t\right)(1-\delta)\lambda_t^{\frac{1}{1-\delta}}e^{-\frac{Y_t}{1-\delta}}-U\left(x,t\right)\delta r\\=&-U\left(x,t\right)\left(G+(1-\delta)\lambda_t^{\frac{1}{1-\delta}}e^{-\frac{Y_t}{1-\delta}}+\delta r\right).
\end{align*}
This completes the proof.
\end{appendices}

\end{document}